\documentstyle[12pt]{article}
\def\a{\alpha}
\def\l{\lambda}
\def\b{\beta}
\def\w{\omega}
\def\ket#1{|{#1}\rangle}  
\def\bra#1{\langle{#1}|}  
\def\norm#1#2{\langle{#1}|{#2}\rangle}   
\def\gov#1#2#3#4{\pmatrix{{#1}&{#2}\cr{#3}&{#4}}}
\def\govi#1#2#3#4{\left[\matrix{{#1}&{#2}\cr{#3}&{#4}}\right]}
\begin{document}
\begin{titlepage}
\rightline{IMSc/92/55}
\rightline{\em 9th Dec 92}
\baselineskip=24pt
\begin{center}
{\large\bf Three Dimensional Chern-Simons Theory as a Theory of
 Knots and Links III : Compact Semi-simple Group}

\vspace{.5cm}

{\bf P. Rama Devi, T.R. Govindarajan and R.K. Kaul} \\
The Institute of Mathematical Sciences, \\
C.I.T.Campus, Taramani, \\
Madras-600 113, INDIA.
\end{center}

\noindent {\bf Abstract}

Chern-Simons field theory based on a compact non-abelian gauge group is studied
as a theory of knots and links in three dimensions. A method to obtain the
invariants for links made from braids of upto four strands is developed. This
generalizes our earlier work on $SU(2)$ Chern-Simons theory.
\vfill
\hrule
\vskip1mm
\noindent{\em email:~rama,trg,kaul@imsc.ernet.in}
\vskip1cm
\end{titlepage}

After Witten's\cite{wit} pioneering work relating Jones polynomial invariants
 for knots in three
dimensions\cite{jon} to quantum field theories, there has been enormous
interest in
such theories\cite{at}. Witten has demonstrated that Jones polynomials are the
expectation values of the Wilson loop-operators in three-dimensional $SU(2)$
Chern-Simons theory with doublet representation of $SU(2)$ living on the knots.
The two-variable generalization\cite{oce} of Jones invariants are obtained in
an $SU(N)$
Chern-Simons theory when $N$-dimensional representations are placed on all the
component knots of a link. These results were demonstrated  by proving
that these invariants satisfy the same Alexander-Conway skein relations as the
respective polynomials. As is well known such skein relations can be
recursively solved to obtain the Jones polynomial and its two variable
generalizations for any arbitrary link. Unfortunately, when other than
fundamental representation of $SU(N)$ is placed on the component knots, the
corresponding generalized Alexander-Conway relations do not admit recursive
solutions in general. For this purpose, we need to develop direct methods to
obtain such invariants. One such attempt for $SU(2)$ Chern-Simons theory was
made in our earlier papers\cite{kg1,kg2}. There we had shown that links related
to four strand
braids can directly be obtained in such a theory. We generalize these results
to $SU(N)$ and other non-abelian compact gauge groups here.
Knot operators for toral knots have also been
obtained \cite{lab} by evaluating the matrix elements of Wilson line operators
in
arbitrary representation of $SU(N)$. This has
allowed Labastida,Llatas and Ramallo to obtain the invariants for
toral knots explicitly. The generalized skein relations for
fundamental representations for $SU(N),SO(N),Sp(2N),SU(m,n)$ and
$OSP(m,n)$ have also been obtained in ref\cite{y}.
Kauffman polnomials correspond to $SO(N)$ Chern-Simons theory\cite{kauf}.

In section 2 we shall present two recursion relations for the topological
invariants in Chern-Simons theory based on a compact nonabelian gauge group
$G$. In section 3, invariants for links obtained as closures of multicoloured
two-strand braids are obtained. In section 4, we shall present representations
of Chern-Simons functional integral over a set of three manifolds with
boundaries
and Wilson lines connecting these boundaries. These can be used to obtain
invariants for many links. In section 5, we present the results of
explicit calculations of the invariants for some links for the $SU(N)$ case
as  illustrations. Concluding remarks are presented in the last section 6.
Appendix contains a brief discussion of the duality matrices for the four-point
correlators of the associated Wess-Zumino conformal field theory  on an $S^2$.

\vspace{1cm}

\noindent {\bf 2. Generalized Alexander-Conway Relations}

For a link $L$ consisting of component knots $C_1,C_2,\ldots C_s$ carrying
representations (colours) $R_1,R_2\ldots R_s$ of the gauge group $G$
respectively,we have the Wilson link operator:
$$
W_{R_1,R_2,\ldots R_s} [L] \ = \ \prod_{i=1}^s tr_{R_i} exp \oint_{C_i} A
\eqno(2.1)
$$
where $A$ is matrix valued connection one-form for the gauge group
$G$. Here the trace is over the representation matrices corresponding to
representations $R_i$. We shall be interested in the functional averages of
these operators in Chern-Simons theory in $S^3$.
$$
V_{R_1R_2\ldots R_s} [L] \ = \ \frac{\int [dA] W_{R_1R_2\ldots R_s}[L]
e^{ikS}}{\int [dA] e^{ikS}}
$$
$$
S \ = \ \frac{k}{4\pi} \int tr (A d A + \frac{2}{3} A^3) \eqno(2.2)
$$
Both the integrand as well as the measure in the functional integrals
are metric independent\cite{kr}. Thus the functional averages (2.2) are link
invariants.

Following our earlier papers\cite{kg1,kg2}, we now obtain recursion relations
relating
these link invariants. For this purpose, let us first consider a link $L_{2m}$
in
$S^3$ as shown in Fig.2.1a. It consists of a room A with two strands entering
from below carrying representations $R$ and $R'$. These
two strands are half twisted $2m$ times and then connected to the
strands leaving the room $A$ from above which carry representations $R, R'$.
This link can be thought of as made from gluing two balls $B_1$ and $B_2$ as
shown in Fig.2.1b along their oppositely oriented boundaries, each an $S^2$.
Like in refs\cite{wit,kg1,kg2}. the Chern-Simons functional integrals over
these two balls
will be represented by vectors $\ket{\psi_{2m}(A)}$ and $\bra{\bar{\psi}_0}$.
These
belong to the Hilbert spaces ${\cal H}$ and $\bar{\cal H}$
 associated with the boundaries of these
two balls respectively. The dimensionality of these two vector spaces is given
by the number of conformal blocks of the $G_k$ Wess-Zumino conformal field
theories on these boundaries, each an $S^2$ with four punctures carrying
representations $R,R',\bar{R},\bar{R}'$. This is given by the number of ways
group singlets can be made from these four representations in accordance with
the fusion rules of the conformal field theory. For sufficiently large $k, R
\otimes R' \ = \ \oplus R_s, s = 0,1,\ldots n$, as allowed by group theory.

Glueing the two balls $B_1$ and $B_2$  can be represented by natural
contraction of the vectors $\ket{\psi_{2m}(A)}$ and $\bra{\bar{\psi}_0}$ :
$$
V_{R,R'} [L_{2m}(A)] \ = \ \norm{\bar{\psi}_0}{ \psi_{2m}(A)} \eqno(2.3)
$$
The vector $\ket{\psi_{2m}(A)}$ is related to the vector $\ket{\psi_0
(A)}$ corresponding to ball $B_1$ with no half-twists in the central two
strands through the half-twist matrix $B(R,R')$ as
$$
B^{2m}_{(R,R')}\ket{\psi_0(A)} \ = \ \ket{\psi_{2m}(A)}
 \eqno(2.4)
$$
The matrix $B$ here introduces right-handed half-twists in parallelly
oriented strands carrying representations $R$ and $R'$. It is an $(n+1) \times
(n+1)$ matrix, where $n+1$ is the number of irreducible representations
in the product $R \otimes R'$. Its eigenvalues are  given by monodromy
properties
of the four-point correlator of primary fields in representations
$R,\bar{R},R',\bar{R'}$ in the Wess-Zumino field theory for group
$G$ on
$S^2$ :
$$
\l^{+}_{s} (R,R') \ = \ (-1)^{\in_s}\exp \ i\pi \left(2h_R +
2h_{R'} + |h_R - h_{R'}|-h_{R_s} \right) \eqno(2.5)
$$
Here $h_R = C_R/(k+C_V)~(C_{R,V}$ are the quadratic Casimir invariants
for the representation  $R$ and adjoint representation
respectively) is
the conformal weight of the primary field in representation $R$
. The factor exp $\pi i (h_R + h_{R'} + |h_R - h_{R'}| )$ here
compensates  the change of framing introduced by a  half unit
twist. The
fore factor $(-)^{\in_s}, \in_s = \pm 1$ counts whether the irreducible
representation $R_s$ in the product $R \otimes R'$ occurs symmetrically or
antisymmetrically. Introducing the variable $q = exp (2\pi i/(k+C_V))$, these
eigen values can be rewritten as
$$
\l^{+}_s (R,R') \ = \ (-1)^{\in_s} q^{C_R + C_{R'}+|C_R -
C_{R'}|/2 - C_{R_s}/2} \eqno(2.6)
$$
Now we write the characteristic equation for the half-twist matrix
$B(R,R')$ as
$$
\sum_{j=\ell}^{n+\ell +1} \alpha_{j-\ell} (B^2)^j \ = \ 0 \eqno(2.7)
$$
where coefficients $\a_0,\a_1, \ldots \a_{n+1}$ are
related to the eigenvalues above. Apply this equation on the vector
$\ket{\psi_0(A)}$ above to obtain
$$
\sum_{j=\ell}^{n+\ell+1} \a_{j-\ell}\ket{\psi_{2j}(A)} \ = \ 0 \eqno(2.8)
$$
We take the inner product with $\bra{\bar{\psi}_0}$ to obtain our first
recursion relation:
\vspace{.5cm}

\noindent {\bf Theorem 1.} \ For the links $L_{2m}(A)$ shown in Fig.2.1a, the
link invariants are related as
$$
\sum^{n+\ell+1}_{j=\ell} \a_{j-\ell} V[L_{2j}(A;R,R')] = 0; \ell = 0,
\pm 1, \pm 2 \ldots \eqno(2.9)
$$
where the coefficients are related to the eigenvalues of the half-twist
matrix $B(R,R')$ above as follows :
\begin{eqnarray*}
\a_0(R,R')&=&(-)^{n+1}\prod_{s=0}
^{n
}\left(\l_s^+(R,R')\right)^2 \\
\a_1(R,R')&=&(-)^n\sum_{s_1\neq s_2\neq {\cdots}
s_n}(\l_{s_1}^+)^2(\l_{s_2}^+)^2\cdots (\l_{s_n}^+)^2\\
&\vdots&\\
\a_{n-1}(R,R')&=&(-)^2\sum_{s_1\neq
s_2}(\l_{s_1}^+)^2(\l_{s_2}^+)^2\\
\a_n(R,R')&=&(-)\sum(\l_s^+)^2\\
\hskip1.5cm\a_{n+1}(R,R')&=&1\hskip10cm (2.10)
\end{eqnarray*}
For links of the type shown in Fig.2.1a where $R$ and $R'$ are the same
representation, a more powerful recursion relation can be obtained:

\vspace{.5cm}
\noindent {\bf Theorem 2.} \ For links $L_m(A;R,R)$ as shown in Fig.2.2.
the link invariants are related as :
$$
\sum^{n+\ell+1}_{j=\ell}\b_{j-\ell}(R,R)V[L_j(A;R,R)] = 0 \eqno(2.11)
$$
where the coefficients are related to the eigenvalues of half-twist
matrix $B(R,R)$ :
\begin{eqnarray*}
\b_0(R,R)&=&(-)^{n+1}\prod_
s \l_s^+(R,R) \\
\b_1(R,R)&=&(-)^n\sum_{s_1\neq s_2\neq {\cdots}
s_n}(\l_{s_1}^+\l_{s_2}^+\cdots \l_{s_n}^+)\\
&\vdots&\\
\b_{n-1}(R,R)&=&(-)^2\sum_{s_1\neq
s_2}(\l_{s_1}^+\l_{s_2}^+)\\
\b_n(R,R)&=&(-)\sum(\l_s^+)\\
\hskip1.5cm\b_{n+1}(R,R)&=&1\hskip10cm (2.12)
\end{eqnarray*}
Here $n+1$ is the number of irreducible representations in the
product, $R \otimes R = \oplus^{n}_{s=0} R_s.$

Next we present the recursion relation amongst the links $L_{2m}(\hat A;R,\bar
R')$ as
shown in Fig.2.3a. These links can again
be thought as obtained by glueing two balls shown in Fig.2.3b. The
functional integrals over these two balls are represented by vectors
$\ket{\chi_{2m}
(\hat{A})}$ and $\bra{\bar{\chi}_0}$ belonging to the Hilbert spaces associated
with
the oppositely oriented boundaries of these balls.
Now here vector $\ket{\chi_{2m}(\hat{A})}$ can be written as
$$
\ket{\chi_{2m}(\hat{A})} \ = \ (\hat{B}(R,{\bar R'}))^{2m}\ket{\chi_0(\hat{A})}
\eqno(2.13)
$$
where the matrix $\hat B(R,\bar R')$ now introduces a half twist in
oppositely oriented strands carrying representations $R,\bar R'$.Its
eigen values in contrast to twist matrix $ B(R,R')$ above are
$$
\l_s^{-}=(-)^{\in_s} q^{-|C_R-C_R'|/2~+~C_{R_s}/2} \eqno(2.14)
$$
where now
$R\otimes R'=\oplus R_s.$
A similar analysis as above then yields us the following
recursion relation:

\noindent {\bf Theorem 3.}~The invariants for links $\hat
L_{2m}(\hat A;R,\bar R')$as shown in Fig 2.3a are related as:
$$
\sum_{j=\ell}^{n+\ell+1}\hat\a_{j-\ell}(R,\bar R')V[\hat L_{2j}(\hat
A;R,\bar R')]=0 \eqno(2.15)
$$
where the coefficients are related to the eigenvalues of
half twist matrix $\hat B(R,\bar R')$ as :
\begin{eqnarray*}
{\hat\a}_0(R,R')&=&(-)^{n+1}\prod_{s=0}
^{n
}\left(\l_s^-(R,\bar R')\right)^2 \\
{\hat\a}_1(R,R')&=&(-)^n\sum_{s_1\neq s_2\neq {\cdots}
s_n}(\l_{s_1}^-)^2(\l_{s_2}^-)^2\cdots (\l_{s_n}^-)^2\\
&\vdots&\\
{\hat\a}_{n-1}(R,R')&=&(-)^2\sum_{s_1\neq
s_2}(\l_{s_1}^-)^2(\l_{s_2}^-)^2\\
{\hat\a}_n(R,R')&=&(-)\sum(\l_s^-)^2\\
\hskip1.5cm{\hat\a}_{n+1}(R,R')&=&1\hskip10cm (2.16)
\end{eqnarray*}
\vskip1cm
\noindent {\bf 3. Invariants for links obtained as closures of two-strand
braids}

Now we shall consider special case of links Figs.2.1 and 2.2 and 2.3. We take
the room A in Fig.2.1a as
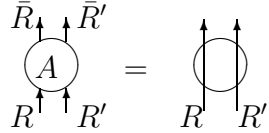
\begin{center}
\begin{picture}(10,7)
\put(5,1.5){\circle{7}}
\put(3,0){$A$}
\put(3.6,4.3){\vector(0,1){3}}
\put(6.8,4.3){\vector(0,1){3}}
\put(6.8,-4.3){\vector(0,1){3}}
\put(3.6,-4.3){\vector(0,1){3}}
\put(0,6){$\bar R$}
\put(8,6){$\bar R'$}
\put(8,-6){$ R'$}
\put(0,-6){$R$}
\end{picture}
{}~~=~~
\begin{picture}(10,7)
\put(5,1.5){\circle{7}}
\put(7,-4){\vector(0,1){11}}
\put(3,-4){\vector(0,1){11}}
\put(7,-6){$R'$}
\put(0,-6){$R$}
\end{picture}
\end{center}
\noindent The functional integral over the ball $B_1$ of Fig.2.1b with room
A as above will be called $\ket{\psi_{2m}(R,R')}$. We expand this vector in
terms of eigen functions of the twist matrix $B(R,R')$ which introduces half
twists
in the central two strands :
$$
B(R,R') \ket{\phi_s} \ = \ \l^{+}_s (R,R')\ket{\phi_s} \eqno(3.1)
$$
and
$$
\ket{\psi_0 (R,R')} \ = \ \sum^n_{s=0} \mu_s \ket{\phi_s} \eqno(3.2a)
$$
$$
\ket{\psi_{2m}(R,R')} \ = \ \sum \mu_s (\l^{+}_s (R,R'))^{2m}\ket{\phi_s}
\eqno(3.2b)
$$
Here we assume $\mu_s$ to be independent of $R,R'$ which we justify
later.
Similarly for the functional integral over the ball $B_2$ of Fig.2.1b.
$$
\bra{\bar{\psi}_0} \ = \ \sum^{n}_{s=0} \mu_s \bra{\phi^s } \eqno(3.2c)
$$
in terms of basis vectors of the dual Hilbert space,$\bar {\cal H}$, which
are also eigen functions of the matrix $B$. The pairing of the basis vectors
for${\cal H}$ and $\bar{\cal H}$ are given by
$\norm{\phi^s}{ \phi_{s'}} = \delta^s_{s'}$. Then the
invariants  for links ${\cal L}(R,R')$ obtained as the closure of two-strand
braids
with $2m$ half-twists as shown in Fig.3.1a are
$$
V [{\cal L}_{2m} (R,R')] \ = \ \sum_s \mu_s \mu_s (\l_s^{+})^{2m} \eqno(3.3)
$$
Now notice for ${\cal L}_0(R,R')$ is two unlinked  unknots carrying
representations $R$ and $R'$.
The invariant for this link is simply the product of  invariants
for two individual unknots, $V_R [U] V_{R'}[U]$. Thus $\sum^{n}_{s=0}
\mu^2_s \ = \ V_R[U] V_{R'}[U]$. Now we use the fact that for two cabled knots
such
as unknots the invariants obey the fusion rules, $V_R[U] V_{R'}[U] = \sum_s
V_{R_s}[U]$. Consistent with this, the invariant for unknot is the
$q$-dimension of the representation living on it: $V_R[U] = dim_q R$. For
example for
the case of $G = SU(N)$, the invariant for the unknot carrying $N$ dimensional
representation is its $q$-dimension = $[N] = (q^{N/2} - q^{- N/2}) / (q^{1/2} -
q^{-
1/2})$.  Thus $\mu_s = \sqrt{dim_q R_s}$ justifying our earlier
statement after Eq.3.2b. This discussion leads us to the following theorem :

\vspace{.5cm}

\noindent {\bf Theorem 4.} \ For links ${\cal L}_{2m}(R,R')$ as shown in
Fig.3.1a
obtained as closure of parallely oriented two-strand braids with
representations
$R,R'$ , the invariants are given by
$$
V [{\cal L}_{2m}(R,R')]~=~\sum_s(dim_qR_s)(\l_s^+(R,R'))^{2m}
\eqno(3.4)
$$
where $R \otimes R' = \oplus R_s$.

Analysing links of the type shown in Figs.2.2 and 2.3a in similar fashion leads
to two similar results.

\vspace{.5cm}

\noindent {\bf Theorem 5.} \ For links ${\cal L}_{m} (R,R)$ as shown in
Fig.3.1b
obtained as closure of parallely oriented braids made of two strands carrying
the same representation $R$, the invariants are
$$
V [{\cal L}_m (R,R)] \ = \ \sum_s (dim_q R_s) (\l_s^{+} (R,R))^m \eqno(3.5)
$$
where $R \otimes R = \oplus R_s$.

\vspace{.5cm}

\noindent {\bf Theorem 6.} \ For links $\hat{\cal L}_{2m} (R,\bar R')$ as shown
in
Fig.3.1c obtained as closure of braids made from two anti-parallely oriented
strands carrying representations $R$ and $\bar R'$ respectively the invariants
are
$$
V [\hat{\cal L}_{2m} (R,\bar R')] \ = \ \sum (dim_qR_s) (\l^{-}_s (R,\bar
R'))^{2m}
\eqno(3.6)
$$
where $R \otimes \bar R' = \oplus R_s$.

After presenting these simple theorems which give us invariants for links made
of closure of two-strand braids, we would like to present functional integrals
on some manifolds with boundaries which can be used to obtain invariants for
more
complicated links.

\vspace{1cm}

\noindent {\bf 4. Some Building Blocks}

In order to develop the method further, we shall write the functional integrals
over some manifold with boundaries carrying four Wilson lines. These can be
used as building blocks to obtain invariants for more complicated links.
Consider the rooms $Q^V_m \gov{.}{.}{R}{R'}
$ and $Q^H_{2p+1} \gov{\bar R'}{\bar{R}}{R}{R'}$ each
a 3- ball with four Wilson lines as indicated in Fig.4.1a. In the first
room $Q^V_m \gov{.}{.}{R}{R'}$, the lower two markings (inlets)
carry representations $R$ and $R'$ respectively and the upper two
markings (outlets) denoted by dots are $\bar{R}$ and $\bar{R'}$ or
$\bar{R}'$ and $\bar{R}$ as $m$ is even or odd. These two rooms have been
redrawn in Fig.4.1b. The functional integral over these two balls have been
called $\ket{\psi(Q^V_m \gov{.}{.}{R}{R'})}$ and $\ket{\psi (Q^H_{2p+1}
\gov{\bar{R}'}{\bar R}{R}{R'})}$ respectively. These can be
constructed from the vectors $\ket{\psi (Q^V_0 \gov{\bar{R}}{\bar{R}'}{R
}{R'})}$ and $\ket{\psi (Q^H_0 \gov{\bar{R}'} {R'}{R}{\bar R})}
 $by applying the half-twist matrix $B$ on the central two strands and
$\hat{B}$ on
the first two side strands respectively. These two vectors from the discussion
in previous section can be expanded in terms of the convenient basis vectors
$\ket{\phi_{R_t} (\bar{R}R R'\bar{R'})}$ which form the eigen functions of
braid
matrix which twists the parallel central two strands (with eigenvalues
$\l^{+}_t (R,,R')$ where $R \otimes R' = \oplus R_{t}$) and $
\ket{\hat{\phi}_{s}(\bar{R'}R R'\bar{R})}$ which are the eigenfunctions of
the braid matrix $\hat{B}$ that twists the first two anti-parallel strands
(with eigen-values $\l^{-}_{R_s} (R,\bar{R}')$ with $R \otimes \bar{R'}
= \oplus R_s)$, respectively :
$$
\ket{\psi (Q^V_0 \gov{\bar{R}}{\bar{R}'}{R}{R'})} \ = \ \sum_t
\sqrt{dim_q R_t} \ket{\phi_{R_t} (\bar{R} R R' \bar{R}')}
$$
$$
\ket{\psi (Q^H_0 \gov{\bar{R}'}{R'}{R}{\bar{R}})} \ = \ \sum_s
\sqrt{dim_q R_s} \ket{\hat\phi_{R_s}(\bar R'R R'\bar R)}\eqno(4.1)
$$

The two bases $\ket{\phi_{R_s}}$ and $\ket{\hat{\phi}_{R_t}}$ are related by
the
duality matrix  of the associated four
point correlators of the corresponding Wess-Zumino model :
$$
\ket{\hat\phi_{R_s}(R_1R_2R_3R_4)}~=~\sum_{R_t}a_{R_sR_t}\left[
\matrix{ R_1 R_2 \cr R_3 R_4} \right] \ket{\phi_{R_t} (R_1 R_2 R_3 R_4)}
\eqno(4.2)
$$
where $R_1 \otimes R_2 \ = \ \oplus R_s, R_2 \otimes R_3 \ = \
\oplus R_t$ and the four representations $R_1, R_2, R_3, R_4$ form a
singlet.  The duality matrix  $a_{R_s R_t} \left[ \matrix {R_1 & R_2 \cr
R_3 & R_4 } \right]$ above,
as in the
$SU(2)_k$ case, are  related to $q$-Racah coefficients for the non
abelian group G.
In  the Appendix we have given the expression for these  in
terms of $q~C-G$  coefficients and listed their other useful properties.

The above discussion can be put together in the form of a theorem :

\noindent{\bf Theorem 7.}The normalized functional integral over the
3-balls for $\ket{\psi(Q^V_m
\pmatrix { . & . \cr R & R'})}$ and $\ket{\psi (Q^H_{2p+1} \pmatrix{ \bar{R}' &
\bar{R} \cr R & R'})}$  as shown in Fig.4.1b can be written in
terms of basis referring to the central two strands as
$$
\ket{\psi (Q^V_{m} \pmatrix{ . & . \cr R & R'})} \ = \ \sum_{R_t} \mu_{R_t}
(Q^V_m \pmatrix {. & . \cr R & R'}) \ \ket{\phi_{R_t} (\cdot R R'\cdot)}
$$
$$
\ket{ \psi (Q^H_{2p+1}( \pmatrix { \bar{R}' & \bar{R} \cr R & R'})} \ = \
\sum_{R_t} \mu_{R_t} (Q^H_{2p+1} \pmatrix{ \bar{R}' & \bar{R} \cr R & R'})
\ket{\phi_{R_t} (\bar{R}'RR' \bar{R}} \eqno(4.3)
$$
where $ R \otimes R' = \oplus R_t$
and
$$
\mu_{R_t} (Q^V_m \pmatrix{ . & . \cr R & R'}) = (\l^{+}_{R_t} (R,
R'))^m \sqrt{dim_q R_t}
$$
$$
\mu_{R_t} (Q^H_{2p+1} \pmatrix{ \bar{R}' & \bar{R} \cr R & R'}) \ = \
\sum_{R_s} (\l^{-}_{R_s} (R,\bar{R}'))^{2p+1} a_{{R_s R_t}} \left[
\matrix{ \bar{R}'& R \cr R'& \bar{R}} \right] \sqrt{dim_q R_s}
\eqno(4.4)$$
where $R \otimes \bar{R}' = \oplus R_s$. The braid matrix eigen
values for a pair of  parallel and anti-parallel strands are given by
Eqns.2.6 and 2.14 respectively.

A similar discussion can be developed for the functional integral over 3-balls
containing rooms $\hat{Q}^V_m \pmatrix { . & . \cr R & \bar{R}'},
\hat{Q}^H_{2p} \pmatrix{ \bar{R}' & R' \cr R & \bar{R} }$ and
$\hat{Q}^{H'}_p \pmatrix{ R' & . \cr R & . }$ as drawn in Fig.4.2a. These
balls have been redrawn in Fig.4.2b. The vectors $\ket{\chi(\hat{Q}^{V}_{m}
\pmatrix{ . & . \cr R & \bar R'})},\ket{\chi(\hat{Q}^{H}_{2p}
\pmatrix{ \bar R' & R' \cr R & \bar R})}$ and $\ket{\chi(\hat{Q}^{H'}_{p}
\pmatrix{ R' & . \cr R & .})}$ respectively represent the functional integral
over these three balls. We present these functional integrals in the form
of a theorem.
\def\r{\rho}

\noindent {\bf Theorem 8.} \ The functional integral $\ket{\hat\chi(Q^V_{2m}
\pmatrix{
. & . \cr R & \bar{R}'})}$, $\ket{\chi(\hat Q^H_{2p} \pmatrix{\bar{R}' & R' \cr
R &
\bar{R}})}$ and $\ket{\chi(\hat Q^{H'}_{p} \pmatrix{{R}' & . \cr R & .})}$ for
the
three balls shown in Fig.4.2b can be expanded in terms of the basis referring
to
the central two strands in each case as :
$$
\ket{\chi(\hat Q_m^V\gov{.}{.}{R}{\bar R'})}~=~\sum_{R_s}{\hat\mu_{R_s}}(\hat
Q_m^V\gov{.}{.}{R}{\bar R'})\ket{\hat\phi_{R_s}(\cdot R \bar R' \cdot)}
$$
$$
\ket{\chi(\hat
Q_{2p}^H\gov{\bar R'}{R'}{R}{\bar R})}~=~\sum_{R_u}{\hat\mu_{R_u}}(\hat
Q_{2p}^H\gov{\bar R'}{R'}{R}{\bar R})\ket{\hat\phi_{R_u}(\bar R' R \bar R
R')}
$$
$$
\ket{\chi(\hat
Q_{2p}^{H'}\gov{R'}{\bar R'}{R}{\bar R})}~=~\sum_{R_u}{\hat\mu_{R_u}}(\hat
Q_{2p}^{H'}\gov{R'}{\bar R'}{R}{\bar R})\ket{\hat\phi_{R_u}(R' R \bar R
\bar R')}
$$
$$
\ket{\chi(\hat
Q_{2p+1}^{H'}\gov{R'}{\bar R}{R}{\bar R'})}~=~\sum_{R_s}{\hat\mu_{R_s}}(\hat
Q_{2p+1}^{H'}\gov{R'}{\bar R}{R}{\bar R'})\ket{\hat\phi_{R_s}(R' R \bar
R' \bar R)}\eqno(4.5)
$$
\noindent where $R \otimes \bar{R}' \ = \ \oplus  R_s , \ R \otimes \bar{R}
\ = \ \oplus R_u,$  and
$$
\hat\mu_{R_s}(\hat
Q^V_m\gov{.}{.}{R}{\bar R'})~=~(\l_{R_s}^-(R,\bar R'))^m\sqrt{dim_qR_s}
$$
$$
\hat\mu_{R_u}(\hat
Q^H_{2p}\gov{\bar R'}{R'}{R}{\bar
R})~=~\sum_{R_s}\sqrt{dim_qR_s}~a_{R_sR_u}\govi{\bar
R'
}{R}{\bar R}{R'
}(\l_{R_s}^-(\bar R'R))^{2p}
$$
$$
\hat\mu_{R_u}(\hat
Q^{H'}_{2p}\gov{R'}{\bar R'}{R}{\bar R})~=~\sum_{R_t}\sqrt{dim_qR_t
}~a_{{R_t}{R_u}}\govi{R'}{R}{\bar R}{\bar R'}(\l_{R_t}^+(R' R))^{2p}
$$
$$
\hat\mu_{R_s}(\hat
Q^{H'}_{2p+1}\gov{R'}{\bar R}{R}{\bar
R'})~=~\sum_{R_t}\sqrt{dim_qR_t}~a_{{R_t}{R_s}}\govi{R'}{R}{\bar R'}{\bar
R}(\l_{R_t}^+(R R'))^{2p+1}\eqno(4.6)
$$
\noindent where $R \otimes R' = \oplus R_t.$

\noindent This generalizes theorem 6 of \cite{kg2}.

Now we may extend our discussion to functional integrals over three manifolds
with more than two boundaries. For example consider the manifold with $r$
boundaries, each boundary an $S^2$ with four punctures and Wilson lines
connecting them as shown in Fig.4.3. The four representations on each of these
boundaries are such that they make a group singlet. Then the functional
integral over this manifold is given by the following theorem:

\vspace{.5cm}

\noindent {\bf Theorem 9.} \ The normalized functional integral for the
manifold with $r$ boundaries as shown in Fig.4.3 in terms of the basis $
\ket{\phi_{R_s}^{(i)}}$ of the Hilbert spaces ${\cal H}^{(i)} , i=1,2,\ldots r$
associated with these boundaries with four punctures each as indicated is given
by
$$
\nu_r \ = \ \sum_{R_s} \frac{\ket {\phi_{R_s}^{(1)} } \ket {\phi^{(2)}_{R_s} }
\cdots
\ket{\phi^r_{R_s} }} {(dim_q R_s)^{\frac{r-2}{2}}} \eqno(4.7)
$$
These bases $| \phi^{(i)}_{R_s}>$ refer to the parallel side-two strands in
each
case. The functional integral is normalised by multiplying  a factor
$(N)^{(r-1)/2}$, where $N$ is the functional
integral over the empty boundaryless manifold $S^3$. The summation
over $R_s$ is over those irreducible representations  which are common to the
products $R_1 \otimes R_2, R_3 \otimes R_4, R_5 \otimes R_6 \ldots R_{2r-1}
\otimes R_{2r}$.

This theorem in the case of manifolds with one and two boundaries as shown in
Figs.4.4 (a) and (b) reduces to
$$
\nu_1 \ = \ \sum_{R_s} \sqrt{dim R_s} \ket{ \phi^{(1)}_{R_s} (R_1R_2\bar{R}_2
\bar{R}_1)} \eqno(4.8)
$$
$$
\nu_2 \ = \ \sum_{R_s}\ket{ \phi_{R_s}^{(1)} (R_1 R_2 \bar{R}_3 \bar{R}_4 )}
\ket{\phi_{R_s}^{(2)} (\bar{R}_1 \bar{R}_2 R_3 R_4)} \eqno(4.9)
$$
The proof of theorem 9 can easily be developed by induction, by
successively glueing the manifold shown in Fig.4.4(a) onto one of the
boundaries
of the manifold shown in Fig.4.3.

\vspace{1cm}

\noindent {\bf 5. Some explicit calculations }

With the building blocks presented above, we are in a position to evaluate the
invariants for all links made of upto four strands. This we shall now
illustrate by
studying some examples. For definiteness let us take the gauge group $G$ to be
$SU(N)$ and place representation given by Youngs tableaux containing $n$ boxes
in a row,
\setlength{\unitlength}{1cm}
$R_n =
\begin{picture}(2,.7)
\put(0,0){\line(1,0){1.8}}
\put(0,.3){\line(1,0){1.8}}
\multiput(0,0)(.3,0){7}{\line(0,1){.3}}
\put(.7,.5){\vector(-1,0){.7}}
\put(1.1,.5){\vector(1,0){.7}}
\put(.8,.5){$n$}
\end{picture}
$
on all the component knots of a link. Then the eigen
values of the braid matrix that introduces right handed half-twist in parallely
oriented strands carrying this representation are (Eqn.2.5) :
$$
\l^{+}_\ell (R_n, R_n) \ = \ (-)^{n-\ell}\w^{n/2} q^{\frac{n(n+1)}{2} -
\frac{\ell(\ell +1)}{2}} ~~;~~ \ell \ = \ 0,1,\ldots n\eqno(5.1)
$$
where we  have introduced a convenient variable $\w = q^{N-1}$. Here
these eigen values for $\ell = 0,1,2,\ldots n$ correspond respectively to the
irreducible representation in the product $R_n \otimes R_n = \oplus_{\ell =0}^n
\r_\ell$ as
$$
\begin{picture}(10,1.5)
\put(0,0){\line(1,0){1.8}}
\put(0,.3){\line(1,0){1.8}}
\multiput(0,0)(.3,0){7}{\line(0,1){.3}}
\put(.7,.5){\vector(-1,0){.7}}
\put(1.1,.5){\vector(1,0){.7}}
\put(.8,.5){$n$}
\put(2,0){$\otimes$}
\put(.8,-.4){$R_n$}
\put(2.6,0){\line(1,0){1.8}}
\put(2.6,.3){\line(1,0){1.8}}
\multiput(2.6,0)(.3,0){7}{\line(0,1){.3}}
\put(3.3,.5){\vector(-1,0){.7}}
\put(3.7,.5){\vector(1,0){.7}}
\put(3.4,.5){$n$}
\put(3.4,-.4){$R_n$}
\put(4.8,0){$=$}
\put(5.4,0){$\oplus_{\ell=0}^n$}
\put(6.5,0){\line(1,0){3}}
\put(6.5,.3){\line(1,0){3}}
\put(6.5,-.3){\line(1,0){1.8}}
\multiput(6.5,0)(.3,0){11}{\line(0,1){.3}}
\multiput(6.5,0)(.3,0){7}{\line(0,-1){.3}}
\put(8,-.6){$\r_\ell$}
\put(6.8,.5){\vector(-1,0){.3}}
\put(8,.5){\vector(1,0){.3}}
\put(6.9,.5){$n-\ell$}
\put(8.6,.5){\vector(-1,0){.3}}
\put(9.2,.5){\vector(1,0){.3}}
\put(8.8,.5){$2\ell$}
\end{picture}
$$

On the other hand, the eigenvalues of the braid matrix that
introduces right-handed half-twists in anti-parallely oriented strands
(Eqn.2.14) is given by
$$
\l^{-}_\ell (R_n, \bar{R}_n) \ = \ (-)^\ell\w^{\ell/2} q^{\ell^2/2}
{}~~;~~\ell = 0,1,\ldots , n\eqno(5.2)
$$
Here these eigenvalues for $\ell = 0,1,2,\ldots n$ correspond
respectively to the irreducible representations in the product $R_n \otimes
\bar{R}_n = \oplus \hat{\rho}_\ell$ :
$$
\begin{picture}(10,1.5)
\put(0,0){\line(1,0){1.8}}
\put(0,.3){\line(1,0){1.8}}
\multiput(0,0)(.3,0){7}{\line(0,1){.3}}
\multiput(.15,.15)(.3,0){6}{$.$}
\put(.7,.5){\vector(-1,0){.7}}
\put(1.1,.5){\vector(1,0){.7}}
\put(.8,.5){$n$}
\put(2,0){$\otimes$}
\put(.8,-.4){$\bar R_n$}
\put(2.6,0){\line(1,0){1.8}}
\put(2.6,.3){\line(1,0){1.8}}
\multiput(2.6,0)(.3,0){7}{\line(0,1){.3}}
\put(3.3,.5){\vector(-1,0){.7}}
\put(3.7,.5){\vector(1,0){.7}}
\put(3.4,.5){$n$}
\put(3.4,-.4){$R_n$}
\put(4.8,0){$=$}
\put(5.4,0){$\oplus_{\ell=0}^n$}
\put(6.5,0){\line(1,0){2.4}}
\put(6.5,.3){\line(1,0){2.4}}
\multiput(6.5,0)(.3,0){9}{\line(0,1){.3}}
\multiput(6.65,.15)(.3,0){4}{$.$}
\put(7.5,-.4){$\hat\r_\ell$}
\put(6.8,.5){\vector(-1,0){.3}}
\put(7.4,.5){\vector(1,0){.3}}
\put(7,.5){$\ell$}
\put(8,.5){\vector(-1,0){.3}}
\put(8.6,.5){\vector(1,0){.3}}
\put(8.2,.5){$\ell$}
\end{picture}
$$

\noindent Here a box with dot represents a column of length $N-1$.

Using these eigenvalues of the braid matrix and the building blocks developed
earlier, we have as illustrations calculated explicitly the invariants for some
knots shown in Fig.5.1. The results are as follows :
\begin{eqnarray*}
0_1: V&=&dim_qR_n\\
3_1:
V&=&\sum^n_{\ell=0}dim_q\r_{\ell}(-)^{n-\ell}\w^{-3n/2}q^{-3(n(n+1)/2-\ell(\ell
+1)/2)}\\
4_1: V&=&
\sum_{\ell,j=0}^n\sqrt{dim_q\hat\r_jdim_q\hat\r_\ell}~a_{\hat\r_j\hat\r_\ell}
\w^{{j\over 2}-{\ell\over2}}q^{{j^2\over 2}-{\ell^2\over 2}}\\
5_1:
V&=&\sum^n_{\ell=0}dim_q\r_{\ell}(-)^{n-\ell}\w^{-5n/2}q^{-5(n(n+1)/2-\ell(\ell
+1)/2)}\\
5_2:
V&=&\sum_{\ell,j=0}^n\sqrt{dim_q\hat\r_jdim_q\r_\ell}~a_{\hat\r_j\bar\r_\ell}
\w^{n+{3j\over 2}}q^{n(n+1)-\ell(\ell+1)+{3j^2\over
2}}\\
6_1:
V&=&\sum_{\ell,j=0}^n\sqrt{dim_q\hat\r_jdim_q\hat\r_\ell}~a_{\hat\r_j\hat\r_\ell}
\w^{-\ell+2j}q^{-\ell^2+2j^2}\\
7_1:
V&=&\sum^n_{\ell=0}dim_q\r_{\ell}(-)^{n-\ell}\w^{-7n/2}q^{-7(n(n+1)/2-\ell(\ell
+1)/2)}\\
7_2:
V&=&\sum_{\ell,j=0}^n\sqrt{dim_q{\hat\r_jdim_q\r_\ell}}~a_{\hat\r_j\bar\r_\ell}
\w^{-n-{5j\over 2}}q^{-n(n+1)+\ell(\ell+1)-{5j^2\over 2}}
\end{eqnarray*}
\noindent Here the duality matrices $a_{\hat{\rho}_i \hat{\rho}_j}
\left[\matrix{
\bar{R}_n & R_n \cr \bar{R}_n & R_n}\right]$ and
$a_{\hat{\rho}_i \bar{\rho}_j} \left[\matrix{{R}_n & \bar R_n \cr
\bar{R}_n & R_n}\right]$ have been written simply as
$a_{\hat{\rho}_i \hat{\rho}_j}$ and $a_{\hat{\rho}_i \bar{\rho}_j}$
respectively.

In particular for $n=1$ and $n=2$~($R_1 =
\begin{picture}(.4,.7)
\put(0,0){\line(1,0){.3}}
\put(0,.3){\line(1,0){.3}}
\multiput(0,0)(.3,0){2}{\line(0,1){.3}}
\end{picture}
$ and $R_2 =
\begin{picture}(.7,.7)
\put(0,0){\line(1,0){.6}}
\put(0,.3){\line(1,0){.6}}
\multiput(0,0)(.3,0){3}{\line(0,1){.3}}
\end{picture}
$) the
duality matrices are explicitly given by:
$$
a_{\hat{\rho}_i \hat{\rho}_j} \ = \ \frac{1}{dim_q R_1} \pmatrix { \sqrt{dim_q
\hat{\rho}_0} & \sqrt{ dim \hat{\rho}_1} \cr \sqrt{dim_q \hat{\rho}_1} &
-\sqrt{
dim_q \hat{\rho}_0}} \eqno(5.5)
$$
and
$$
a_{\hat{\rho}_i \hat{\rho}_j} \ = \ \frac{1}{dim_q R_2} \pmatrix { \sqrt{dim_q
\hat{\rho}_0} & \sqrt{ dim_q \hat{\rho}_1} & \sqrt{dim_q \hat{\rho}_2} \cr
\sqrt{dim_q \hat{\rho}_1} & \frac{ dim_q \hat{\rho}_2}{dim_q R_2-1} -1 &
\frac{-\sqrt{dim_q \hat{\rho}_1 dim_q \hat{\rho}_2}}{dim_q R_2-1} \cr
\sqrt{dim_q \hat{\rho}_2} & - \frac{\sqrt{dim_q \hat{\rho}_1 dim_q
\hat{\rho}_2}}{dim_q R_2 -1} & \frac{1-dim_q \hat\r_2 + dim_q
\hat{\rho}_1}{dim_q
R_2 - 1} } \eqno(5.6)
$$
The other duality matrix $a_{\hat{\rho}_i \bar{\rho}_j}$ can also be written
down exploiting the "orthoganality" and symmetry properties given in the
Appendix.

\vspace{1cm}

\noindent {\bf 6. Conclusions}

Here we have presented a method to obtain the expectation values of Wilson link
operators for links made up of upto four strands in a Chern-Simons theory on
$S^3$, based on an arbitrary compact non-abelian gauge group $G$. This
generalises our discussion based on $SU(2)$ presented in refs\cite{kg1,kg2}.
A whole variety of new link invariants are thus obtained. As a special case for
gauge group $SU(N)$ with the $N$ dimensional fundamental representation
$\begin{picture}(.4,.7)
\put(0,0){\line(1,0){.3}}
\put(0,.3){\line(1,0){.3}}
\multiput(0,0)(.3,0){2}{\line(0,1){.3}}
\end{picture}
$
placed on all the component knots yields the two variable generalization
of Jones
polynomial, the so called HOMFLY invariant. Another special case when the
representation
$ \begin{picture}(.7,.7)
\put(0,0){\line(1,0){.6}}
\put(0,.3){\line(1,0){.6}}
\multiput(0,0)(.3,0){3}{\line(0,1){.3}}
\end{picture}
$ of $SU(N)$ is placed on the component knots leads to the
two variable generalized polynomial of Akutsu,Deguchi and
Wadati\cite{adw}. While this method can be
used to obtain invariants for all links which can be made from upto four
strands, the others still stay elusive.
\newpage
\noindent{\bf Appendix}
\vskip1cm
The duality matrix $a_{R^{(s)} R^{(t)}} \left[ \matrix{ R_1 & R_2 \cr R_3 &
R_4} \right]$ introduced in the text tells us how the conformal blocks for
four-point correlators of primary fields in representations $R_1 R_2 R_3 R_4$
in
$SU(N)_k$ Wess-Zumino conformal field theory transform into each other under
duality transformation :
\setlength{\unitlength}{1cm}
$$
\begin{picture}(11,3)
\put(1,0){\vector(1,0){1}}
\put(2,0){\line(1,0){1}}
\put(1,0){\line(0,1){2}}
\put(1,2){\line(1,1){.7}}
\put(0.3,2.7){\line(1,-1){.7}}
\put(3,0){\line(0,1){2}}
\put(3,2){\line(1,1){.7}}
\put(2.3,2.7){\line(1,-1){.7}}
\put(4,.5){\vector(1,0){4}}
\put(10,2){\vector(0,-1){1}}
\put(10,1){\line(0,-1){1}}
\put(10,2){\line(1,1){.7}}
\put(9.3,2.7){\line(1,-1){.7}}
\put(10,0){\line(1,1){2.7}}
\put(7.3,2.7){\line(1,-1){2.7}}
\put(4.5,1.25){$a_{R^{(s)}R^{(t)}}\left[\matrix{R_1&R_2\cr R_3&R_4}\right]$}
\put(.3,2.9){$R_1$}
\put(1.7,2.9){$R_2$}
\put(1.1,.4){$R^{(s)}$}
\put(3.1,.4){$\bar R^{(s)}$}
\put(3.7,2.9){$R_4$}
\put(2.3,2.9){$R_3$}
\put(9.3,.8){$\bar R^{(t)}$}
\put(10.1,1.7){$R^{(t)}$}
\put(7.3,2.9){$R_1$}
\put(9.3,2.9){$R_2$}
\put(10.7,2.9){$R_3$}
\put(12.7,2.9){$R_4$}
\end{picture}
\eqno(A.1)
$$

\noindent Here $R^{(s)}$ are the irreducible representation in $R_1 \otimes
R_2$ such that $\bar{R}^{(s)}$ is contained in $R_3 \otimes R_4$. And also
$R^{(t)}$ are contained in $R_2 \otimes R_3$ such that $\bar{R}^{(t)}$ are
in $R_1 \otimes R_4$. These duality matrices are given by the $SU(N)_q$ Racah
coefficients. These relate two ways of combining the four representation $R_1,
R_2, R_3, R_4$ of $SU(N)_q$ into singlets. For example we could first combine
$R_1$
and $R_2$ into $R_{12}^{(s)}$. This combined with $R_3$ should yield
$\bar{R}_4$ which then with
fourth representation $R_4$ will give a singlet. Alternatively we could first
combine $R_2$ and $R_3$  to give $R^{(t)}_{23}$. This with $R_1$ then
would have to give $\bar{R}_4$ which with $R_4$ would give a singlet. Calling
these
vectors $\ket{\phi^{(s)}}$ and $\ket{\phi'^{(t)}}$ in (A.1) above respectively,
we can write them
in terms of the representation vectors
$\ket{ R_1 \mu_1}, \ket{ R_2, \mu_2}, \ket{ R_3 \mu_3}, \ket{ R_4 \mu_4} $
(where $\mu_i$ are corresponding weights) through $q-CG$ coefficients
of $SU_q(N)$ :
$$
\ket{ \phi^{(s)}} \ = \ \sum_{\mu_1, \mu_2, \mu_3, \mu_4,\mu_{12}^{(s)}}
C^{R^{(s)}_{12} \mu_{12}^{(s)}}_{R_1 \mu_1 R_2 \mu_2}  C^{\bar{R}_4 -
\mu_4}_{R_{12}^{(s)} \mu_{12}^{(s)} R_{3} \mu_3} C^{00}_{\bar{R}_4 -
\mu_4 R_4 \mu_4} \ket{R_1 \mu_1} \ket{ R_2 \mu_2} \ket{R_3 \mu_3} \ket{R_4
\mu_4} \eqno(A.1)
$$
$$
\ket{\phi'^{(t)}} \ = \ {\sum_{\l_1, \l_2, \l_3,
\l_4,\l_{23}^{(t)}}} C^{R^{(t)}_{23} \l_{23}^{(t)}}_{R_2 \l_2 R_3
\l_3} C^{\bar{R}_4 -\l_4}_{R_1 \l_1 R_{23}^{(t)}
\l^{(t)}_{23}}C^{00}_{\bar{R}_4 - \l_4 R_4 \l_4} \ket{R_1 \l_1}
\ket{ R_2 \l_2} \ket{R_3 \l_3} \ket{R_4 \l_4} \eqno(A.2)
$$
Here C's are the $q-CG$ coefficients. Now from these
$$
a_{R^{(s)} R^{(t)}} \left[ \matrix{ R_1 & R_2 \cr R_3 & R_4} \right] \equiv \
\norm{\phi'^{(t)}} {\phi^{(s)}}
$$
$$
= \sum_{fix \mu_4} C^{R^{(s)}_{12} \mu^{(s)}_{12}}_{R_1 \mu_1 R_2 \mu_2}
C^{\bar{R}_4 - \mu_4}_{R^{(s)}_{12} \mu^{(s)}_{12} R_3 \mu_3} C^{R^{(t)}_{23}
\mu^{(t)}_{23}}_{R_2 \mu_2 R_3 \mu_3} C^{\bar{R}_4 - \mu_4}_{R_1 \mu_1
R_{23}^{(t)} \mu_{23}^{(t)}} \eqno(A.3)
$$
These duality matrices satisfy the following relations :
$$
a_{R^{(s)}_{12} R^{(t)}_{23}} \left[ \matrix{ R_1 & R_2 \cr R_3 & R_4} \right]
\ = \ a_{\bar{R}_{12}^{(s)} R^{(t)}_{23}} \left[ \matrix{ R_3 & R_4 \cr R_1 &
R_2} \right]
$$
$$
= \ a_{R^{(t)}_{23} R^{(s)}_{12}} \left[ \matrix { R_3 & R_2 \cr R_1 & R_4}
\right] = a_{\bar{R}^{(t)}_{23} \bar{R}_{12}^{(s)}} \left[ \matrix{ R_1 & R_4
\cr R_3 & R_2} \right]
$$
and the ``orthogonality'' relation.
$$
\sum_{R^{(t)}_{23}} a_{R_{12}^{(s)} R^{(t)}_{23}} \left[ \matrix{ R_1 & R_2 \cr
R_3 & R_4} \right] a_{R^{(s')}_{12} R^{(t)}_{23}} \left[ \matrix{ R_1 & R_2 \cr
R_3 & R_4} \right] \ = \ \delta_{R_{12}^{(s)} R^{(s')}_{12}} \eqno(A.5)
$$

\newpage
\flushleft{\bf Figure Captions:}\\
\begin{itemize}
\item[Fig.2.1]  Composition of two balls $B_{1}\,, B_{2}$ leads
to the multicoloured (with representations $R$ and $R'$) link
$L_{2m}(A;R,R')$ in $S^3$.
\item[Fig.2.2]  Composition of two balls $B_{1}\,, B_{2}$ leads
to the knot/link $L_{m}(A;R,R)$ in $S^3$.
\item[Fig.2.3] Composition of two balls $B_{1}\,, B_{2}$ leads
to the multicoloured link $\hat{L}_{2m}(\hat{A};R,\bar R')$ in $S^3$.
\item[Fig.3.1]  Closure of (a,b) parallely and (c)
antiparallely oriented two-strand braids with 2m,m,2m half-twists, ${\cal
L}_{2m}
(R,R')$,${\cal L}_m(R,R)$ and $\hat{\cal L}_{2m}(R,\bar R')$ respectively.
\item[Fig.4.1] (a) Rooms $Q^{V}_{m} \pmatrix{ . & .
\cr R & R'}$ and $Q^H_{2p+1} \pmatrix{\bar R' & \bar R \cr R & R'}$
and \\(b) corresponding states $\ket{\psi (Q^V_m \pmatrix{ . & . \cr R &
R'}}$ and $\ket{\psi (Q^H_{2p+1} \pmatrix{\bar R' & \bar R \cr R& R'})}$
\item[Fig.4.2] (a) Rooms $\hat{Q}^V_m \pmatrix{ . & .
\cr R & \bar R'},\hat{Q}^H_{2p} \pmatrix{\bar R'& R'\cr R & \bar R}$ and
$\hat{Q}^{H'}_{p} \pmatrix{R'& . \cr R & .}$ and \\(b) corresponding
states $\ket{\chi (Q^V_m \pmatrix{ . & . \cr R & \bar R'})}$,
$\ket{\chi(\hat{Q}^H_{2p} \pmatrix{\bar R'&R'\cr R& \bar R})}$ and
$\ket{\chi(\hat{Q}^{H'}_{p} \pmatrix{R'& . \cr R & .})}$
\item[Fig.4.3] Functional integrals over a manifold
with $r$ boundaries (with orientation $\in = +$) $\nu_r$ of Eqn.4.7
\item[Fig.4.4] Diagrammatic representation of
functional integrals $\nu_1$ and $\nu_2$.
\item[Fig.5.1] Some knot projections upto seven crossings.
\end{itemize}

\end{document}